\documentclass[11pt]{article}

\usepackage{a4,graphicx,epsfig,rotating}

\begin{document}
\title{WIMP direct detection overview\footnote{Invited review given at
NEUTRINO 2002, Munich, Germany, May 25-30, 2002}}

\author{Y. Ramachers\footnote{e-mail: yorck.ramachers@lngs.infn.it}\\
        Oxford University, Denys Wilkinson Building, \\
Physics Department, OX1 3RH Oxford, UK} 
       
\date{}
\maketitle
\begin{abstract}
This review on weakly interacting massive particle (WIMP) dark matter
direct detection focuses on experimental approaches and the corresponding
physics basics. The presentation is intended to
provide a quick and concise introduction for non-specialists to this
fast evolving topic of astroparticle physics. 
\vspace{1pc}
\end{abstract}

\section{Introduction}
There exists a large collection of measurements providing convincing
evidence in favor of the existence of dark matter in the universe 
(reviewed by L. Bergstroem at this conference, see \cite{lars}
and references therein). The nature of dark matter in the universe is
among the most demanding questions in astroparticle physics. Dark
matter refers to matter which is inferred astronomically only through
its gravitational effects, and which neither absorbs nor emits
sufficient electromagnetic radiation. Moreover, it is likely that the
determination of its nature will yield new information in particle
physics, since there is strong evidence that the dark matter is not
composed of baryons but is rather in some exotic form
\cite{reviews}. One group of main candidates can be classified as weakly
interacting massive particles (WIMPs). A particular candidate favoured
from particle physics is the lightest supersymmetric particle (LSP),
in most modern supersymmetric theories the neutralino. However, WIMP
searches are not specialized to detect the neutralino but any particle
with similar generic properties like a mass above a few GeV and weakly
interacting with normal matter.
\par
Currently many experiments try to reveal the existence and properties
of WIMPs by their direct detection \cite{ddrev} (for indirect
detection approaches, 
see \cite{lars}). Direct detection means to detect WIMP energy
deposition by their elastic scattering off nuclei of specially
designed low background detectors. Since WIMPs are assumed to compose
the major part of the dark halo of our galaxy, the kinematical
constraints determine the general requirements of such a detection
technique. The details of WIMP direct detection will be outlined in
the next section. 
\par
The current experimental status may be briefly summarized as follows:
no WIMPs have been found so far - the first evidence for a
WIMP detection from the DAMA collaboration \cite{dama} still lacks an
independent confirmation of their result. The currently closest
competitors to DAMA, the CDMS collaboration \cite{cdms}, EDELWEISS
\cite{edelweiss} and ZEPLIN \cite{zeplin} are so far not sensitive
enough in order to fully test the announced WIMP-nucleon cross section
versus WIMP--mass region by the DAMA collaboration. Even the most
recent result by the EDELWEISS collaboration, first presented here at
this conference, does not completely test the DAMA region taking into
account all necessary assumptions for such a comparison (see below and
the EDELWEISS presentation at this conference
\cite{edelweiss2}). Finally, many new or upgraded experiments will
soon reach a significantly improved sensitivity level for WIMP
detection as will be outlined in this review. Chances are good to see
an exciting year 2002/3 for the direct detection of dark matter
particles. 
\section{WIMP direct detection physics concepts}
From an experimentalists point of view, one might summarize the main
characteristics of a direct detection experiment with three
key-points. 
\begin{description}
\item[Energy threshold] as low as possible. Since the WIMP signal is
expected to originate from elastic scattering, a featureless, quasi
exponentially decreasing energy spectrum will result. The relevant
energy region will be typically below 100 keV. Therefore, the lower
the energy threshold, the more of the signal can be detected.
\item[Target mass] as high as possible. Since WIMP direct detection
means a rare event search with total rates constrained by experiments
to be roughly below 1 event per kg detector mass per day, one would
need target masses generally above the kilogram scale in order to gain
a sufficient statistic in a reasonable life-time of the experiment.
\item[Background] as low as possible. Two keynotes are worth to
remember here as this is the major parameter for current WIMP direct
searches. First, the signal is a nuclear recoil, i.e. a WIMP scatters
elastically off a nucleus of the detector material thereby producing a
nuclear recoil, which then deposits its energy in the
detector. The operation of WIMP detectors in an
underground laboratory using all the typical precautions of rare event
searches (material selection, shielding) is mandatory. For nuclear
recoil events a generic background contribution originates from 
neutrons. Second, the majority of background consists of electron
recoils from photons (x-ray or gamma-ray radiation) or electrons
(beta radiation). Any means to discriminate between these two types of
recoil energy depositions automatically reduces the background
significantly. 
\end{description}
\par
\begin{figure*}[th]
\epsfxsize=7.8cm
\epsfbox{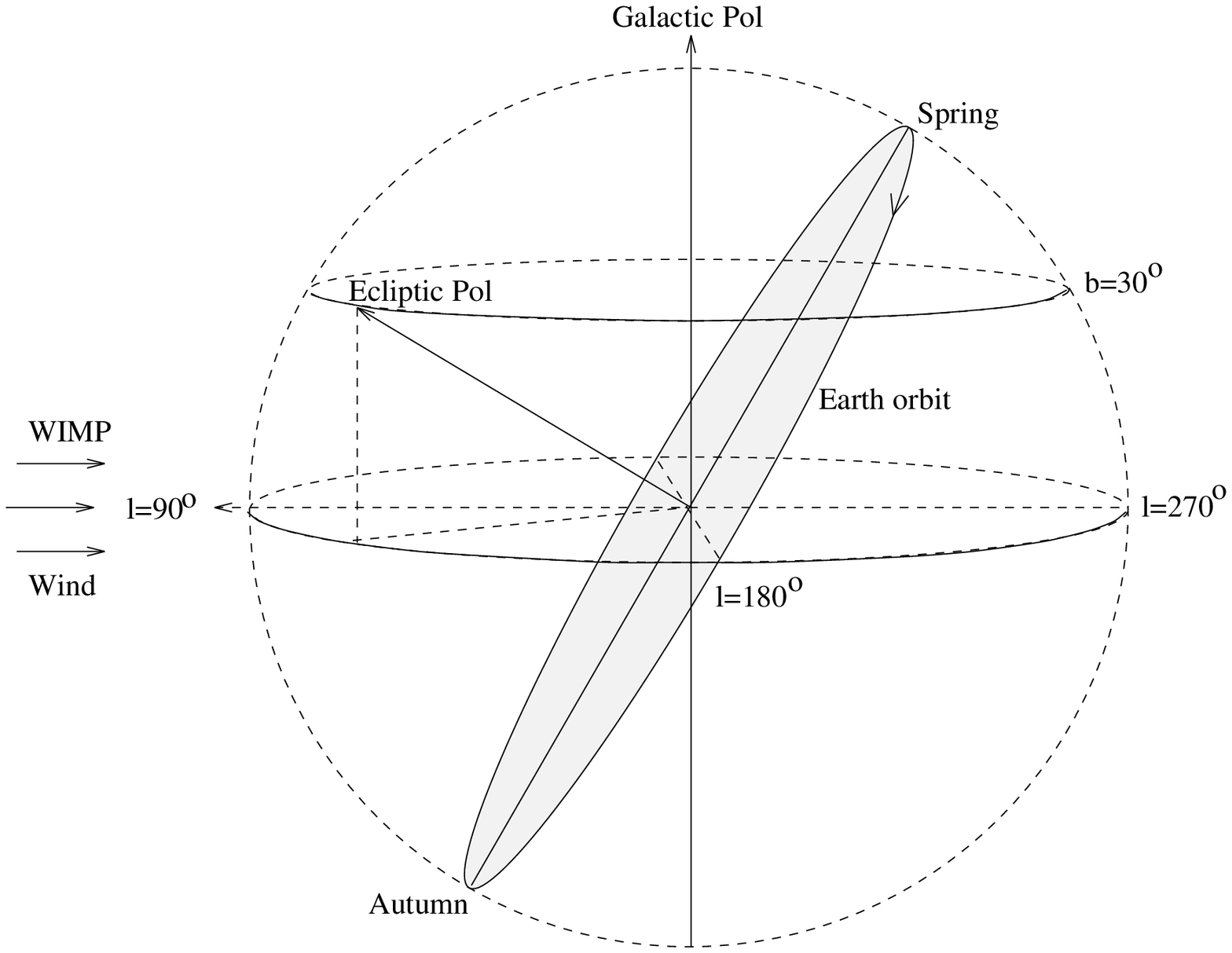}
\hspace{1mm}
\includegraphics[width=5.5cm]{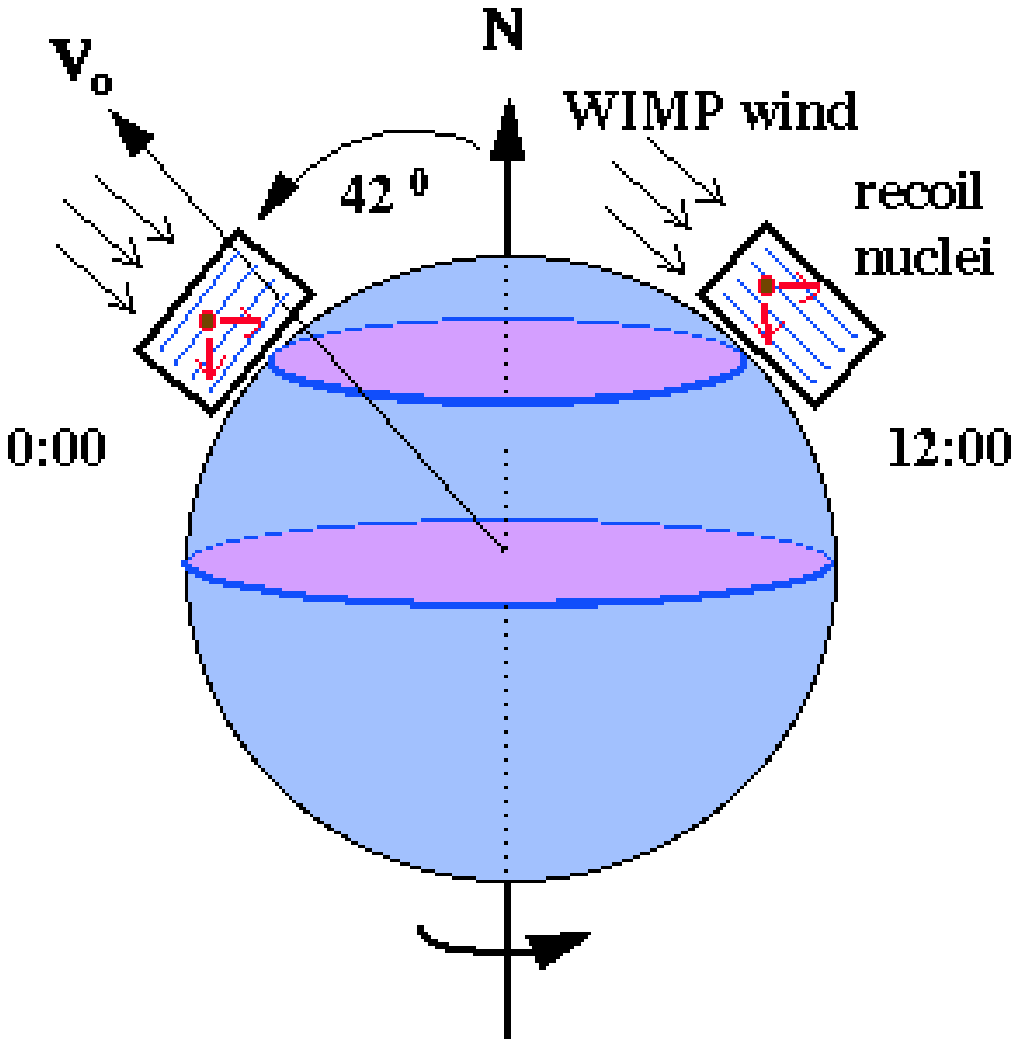}
\caption{The time--dependent WIMP direct detection signatures. The drawing on
the left side displays the Earth orbit in galactic coordinates and the
sun moves to the left with about 220 km/s, inducing a WIMP wind. The
kinetic energy changes in summer compared to winter induce an annual
modulation of count rates. The same WIMP wind induces additionally a
strong asymmetry of nuclear recoil directions which would modulate on
a daily basis as shown in the drawing on the right. The third
signature, not displayed, is the target material dependence of the
rate equation.}
\label{fig2}
\end{figure*}
\par
Most of the physics of WIMP direct detection can be described in
detail by the WIMP-nucleus interaction rate equation
(\ref{eq1}), calculating expected counts per recoil energy, see also
\cite{jungman}: 
\begin{equation}\label{eq1}
\frac{dR}{dQ} =
2\,N_{T}\,\frac{n_{0}\,\sigma_{0}}{m_{w}\,r}\,F^{2}(Q)\,
\int^{v_{max}}_{v_{min}} 
\;\frac{f(v)}{v}\,dv 
\end{equation}
\begin{equation}
v_{min} = \sqrt{\frac{2\,E_{R}}{m_{w}\,r}}\;,\;r =
4\,\frac{m_{w}\,m_{n}}{(m_{w}+m_{n})^{2}}\;.
\end{equation}
This equation can be decomposed into various
contributions from different fields of research, notably particle- and
astro-physics. All numbers or functions described as belonging to
''detector-physics'' in
Tab.~\ref{fig1} are assumed to be well known or possess minor
uncertainties. Most of them are under control of the experimenter like
the amount of target nuclei, N$_{T}$, target nucleus mass, m$_{n}$ and
recoil energy E$_{R}$. An important point to note
about the recoil energy value is that for some types of detection
techniques, see below, the measured energy value does not correspond
to the deposited recoil energy. A detector--specific quenching factor, 
to be calibrated before, has to correct effects of ionisation losses
of nuclear recoil events compared to electron recoil events. 
\par
\begin{table}[htb]
\begin{center}
\caption{Decomposition of the rate equation (\protect{\ref{eq1}}).}
\begin{tabular}{ccc}
Particle- &Astro- &Detector- \\
 Physics &Physics&Physics \\\hline
$m_{w}$ & $n_{0}=\rho_{0}/m_{w}$ & F$^{2}(Q)$ \\
 $\sigma_{0}$ & f(v) & $m_{n}$ \\
   & $v_{max}$ & N$_{T}$ \\
  & & E$_{R}$\\
 unknown &estimates &minor uncertainty \\
\label{fig1}
\end{tabular}
\end{center}
\end{table}
\par
\begin{figure*}[th]
\epsfxsize=10cm
\centering\epsfbox{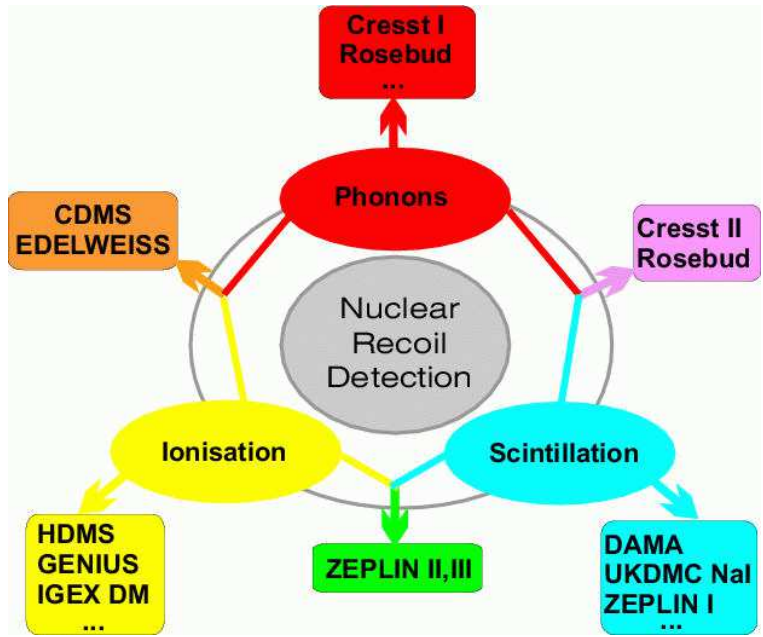}
\caption{Classification scheme for WIMP direct detection techniques.}
\label{fig3}
\end{figure*}
\par
The form factor as a function
of recoil energy, F$^{2}$(Q), has to be calculated specifically for a
given target 
nucleus. It parametrises the loss of coherence of a WIMP interaction with
a nucleus being an extended object. The form factor is an input from
nuclear physics and a detailed treatment of it can be found in
\cite{lewin,jungman}. 
\par
\begin{figure*}[th]
\epsfxsize=5cm
\epsfbox{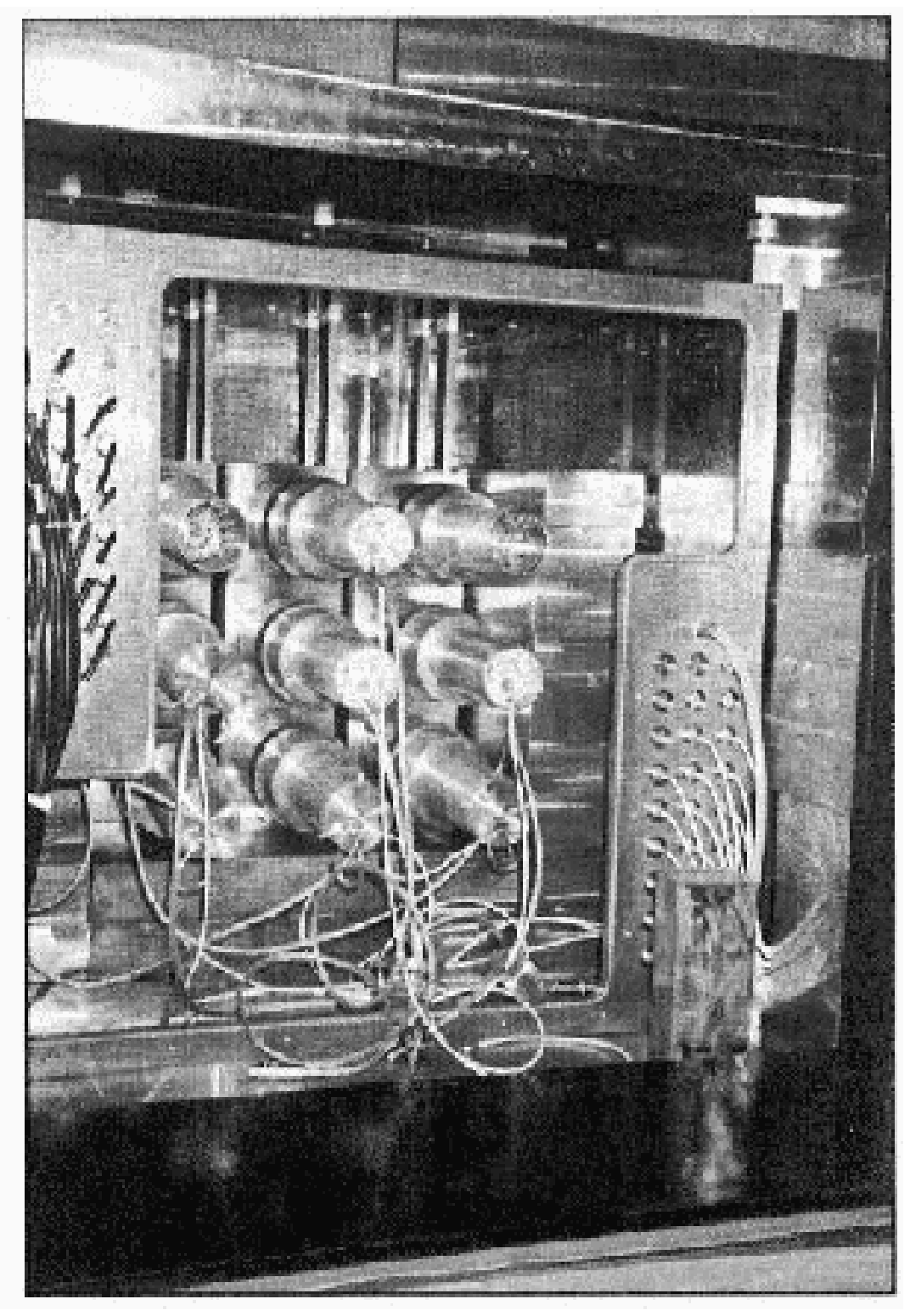}\hspace{1mm}
\epsfxsize=8cm
\epsfbox{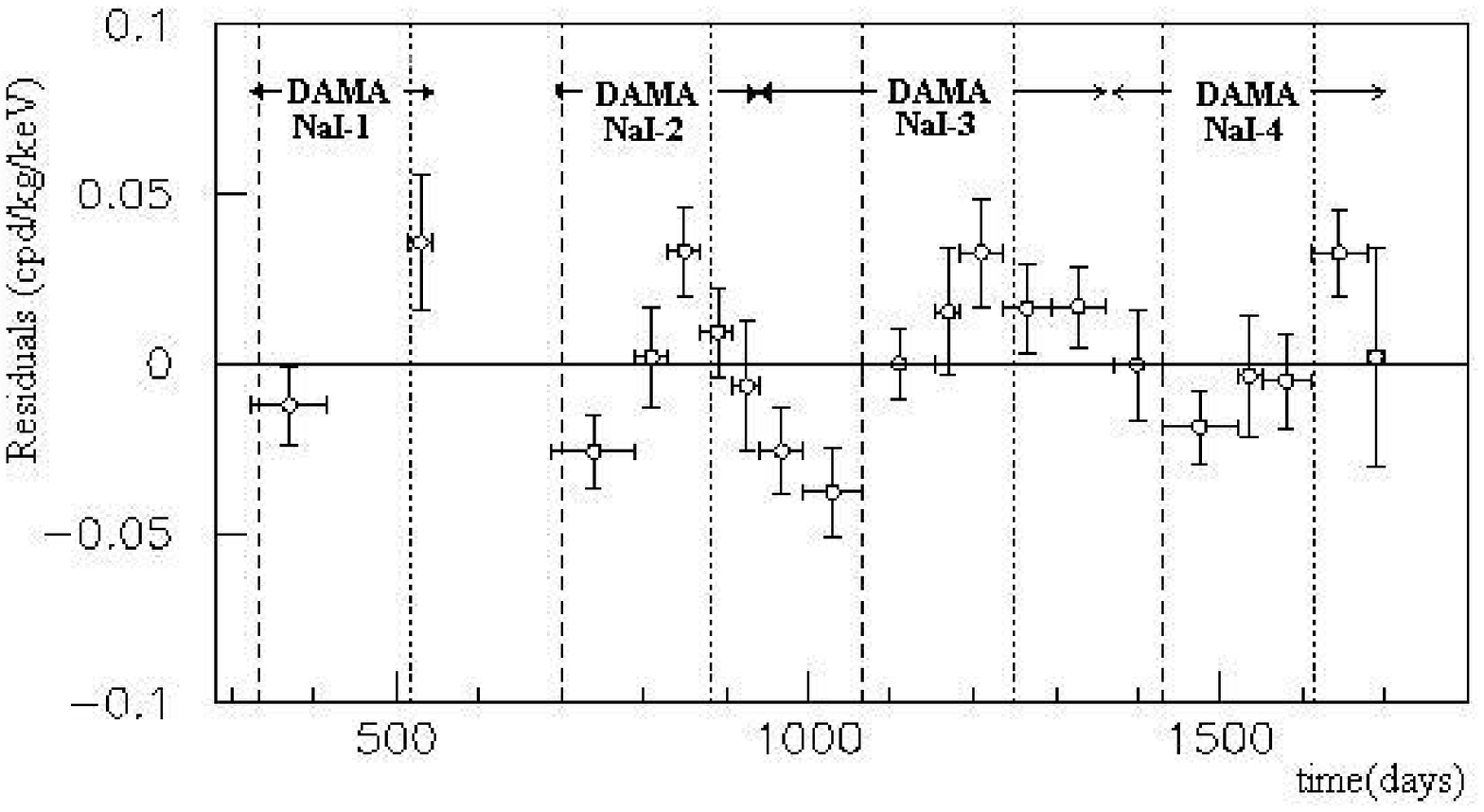}
\caption{An illustration of the DAMA experiment and its announced
evidence. The left panel shows a picture of the experimental setup -
almost 100 kg of NaI scintillation detectors in their special low
background shielding. The right panel shows the residual count rates
as function of time for four years of measurement, displaying
their annual modulation.}
\label{dama}
\end{figure*}
\par
The terms labelled ''astrophysics'' represent values and functions
which are input from astronomy. These describe the source or location
of WIMPs,
i.e. the WIMP dark halo of our galaxy. The escape velocity, v$_{max}$,
determines the  
cutoff of the WIMP energy spectrum at high energies, typically below
100 keV, since  
its value being around 600 km/s. A more important parameter is the
local halo density,  
$\rho_{0}$, since the WIMP signal is directly proportional to its
value. This particular number represents rather an  
assumption than a reasonable guess. It is very uncertain, to a factor
two or even more, since its value is strongly linked to the assumed halo
model. The precise dark halo model would determine also the WIMP
velocity distribution, f(v). However, since any dark halo model is so
far merely an assumption, both, the WIMP halo density and the WIMP
velocity distribution represent a significant systematic
uncertainty. The determination of a dark halo model is currently a
very active topic of research in astrophysics (\cite{white} and
references therein). Meanwhile,
the WIMP direct detection community uses a canonically assumed halo
model, the simplest reasonable model fixing the density at a value of
0.3 GeV/cm$^{3}$ and assumes a Maxwellian velocity distribution for
WIMPs.  
\par
\begin{figure}[th]
\begin{rotate}{270}
\epsfxsize=6cm
\epsfbox{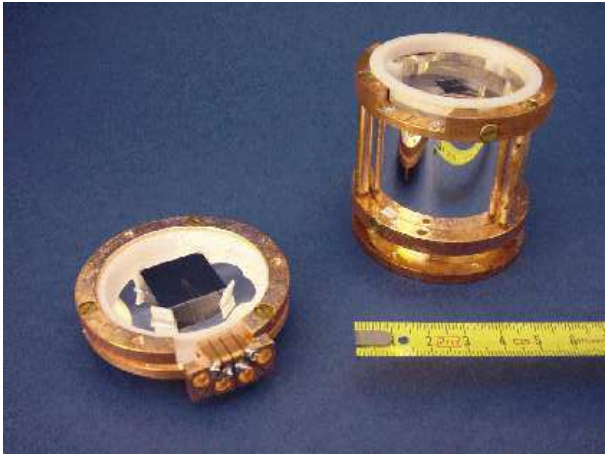}
\end{rotate}
\vspace{6cm}
\caption{The new CRESST II detector module with a CaWO$_{4}$
scintillating crystal. The thin $3\times{}3$ cm$^{2}$ area light detector is
mounted in the cover on the left side of the picture.}
\label{fig4}
\end{figure}
\par
The remaining two numbers in the rate equation are attributed to
particle physics and are completely unknown. They characterize
properties of the unknown WIMP, i.e. its elastic scattering cross
section, $\sigma_{0}$, and its mass, m$_{w}$. These are the numbers to
be determined by direct detection experiments. Consequently, results
of such experiments are given on the $\sigma_{0}$--m$_{w}$ plane. In
fact, there exist two independent representations of results since in
general a non-relativistic WIMP (moving in the galactic gravitational
potential means to move non-relativistic for particle masses under
consideration here) can couple to normal matter
either coherently (scalar coupling) or to the spin of a
nucleus (axial coupling) or both \cite{lewin,jungman,tovey}. The
coherent channel has been the most attractive so far since cross sections
scale as nucleus mass squared (coupling to all nucleons
equally). Therefore, for this spin-independent channel targets made of
high mass nuclei are favoured. Besides the elastic scattering process,
also inelastic scattering has been examined (for a review
\cite{jungman}) but this is beyond the scope of this article. 
\par
Direct consequences of the rate equation are the three WIMP
signatures listed in the caption of Fig.~\ref{fig2}. Two of them
result from a
time-dependence in the rate equation. The annual modulation signature
originates from the revolution of the Earth around the Sun, as 
sketched in Fig.~\ref{fig2}, left panel. This figure shows the Earth
orbit in galactic coordinates, with the sun moving to the left
according to the local rotation curve with about 220 km/s. The
additional velocity (about 15 km/s) of the Earth revolution 
adds in summer and subtracts in winter \cite{freese}, inducing mean
kinetic energy changes for impinging WIMPs. The resulting count rate
modulations as function of recoil energy are claimed to have been
detected by the DAMA experiment, and consequently they announced
evidence for WIMP detection \cite{dama}. No competing background
process mimicking this signature has been identified so far.  
\par
The second time-dependent signature, the diurnal signature
\cite{spergel}, can be 
exploited from detectors which are capable to measure the nuclear
recoil direction. On average, this directionality of WIMP scattering
events should be highly asymmetric with the majority of events
pointing in the 'downwind' direction, i.e. anti-parallel to the Sun
velocity vector. Due to the rotation of the Earth, this
asymmetry would therefore be time-dependent, inducing a diurnal
modulation of events.  
\par
Finally, as the third signature, the rate equation possesses a
complicated dependence on the 
target material, i.e. on the target nucleus mass m$_{n}$. In case
one could achieve sufficiently similar background conditions for two
or more detectors which consist of different target materials, one
might have the chance to measure the characteristic signal ratios,
predicted by the rate equation. So far, no specific proposal exists to
target this particular WIMP signature, however, it might turn out that
it could be successfully applied by the cryogenic detector technology 
(see below and for the signature \cite{smith}). 
\section{WIMP direct detection techniques}
The three minimal requirements of low threshold, reasonably high
target mass and ultra-low background for WIMP direct detection
experiments seem to constrain detector technology quite
significantly. Nevertheless, a large variety of ingeniously designed
detectors currently search for WIMP dark matter or will start
measurements soon. An almost complete classification scheme can be
inspected in Fig.~\ref{fig3}. Note that although it is meant to be as
general as possible, at least four experiments or experiment proposals
do not quite fit into this scheme. These will be mentioned at the end
of this section. 
\par
Three general detection principles for nuclear recoil energy
depositions are shown in Fig.~\ref{fig3}: ionisation, scintillation
and phonon detection, where phonon detection symbolizes the cryogenic
detector technology. Among these three, only scintillation detectors
offer an intrinsic nuclear recoil discrimination mechanism by pulse
shape analysis. This is one of the main reasons for these detectors to
be among the most sensitive experiments for WIMP searches, notably the
DAMA, UKDMC NaI \cite{naiad} and the ZEPLIN I liquid Xenon experiment.
\par
\begin{figure*}[th]
\epsfxsize=9cm
\epsfbox{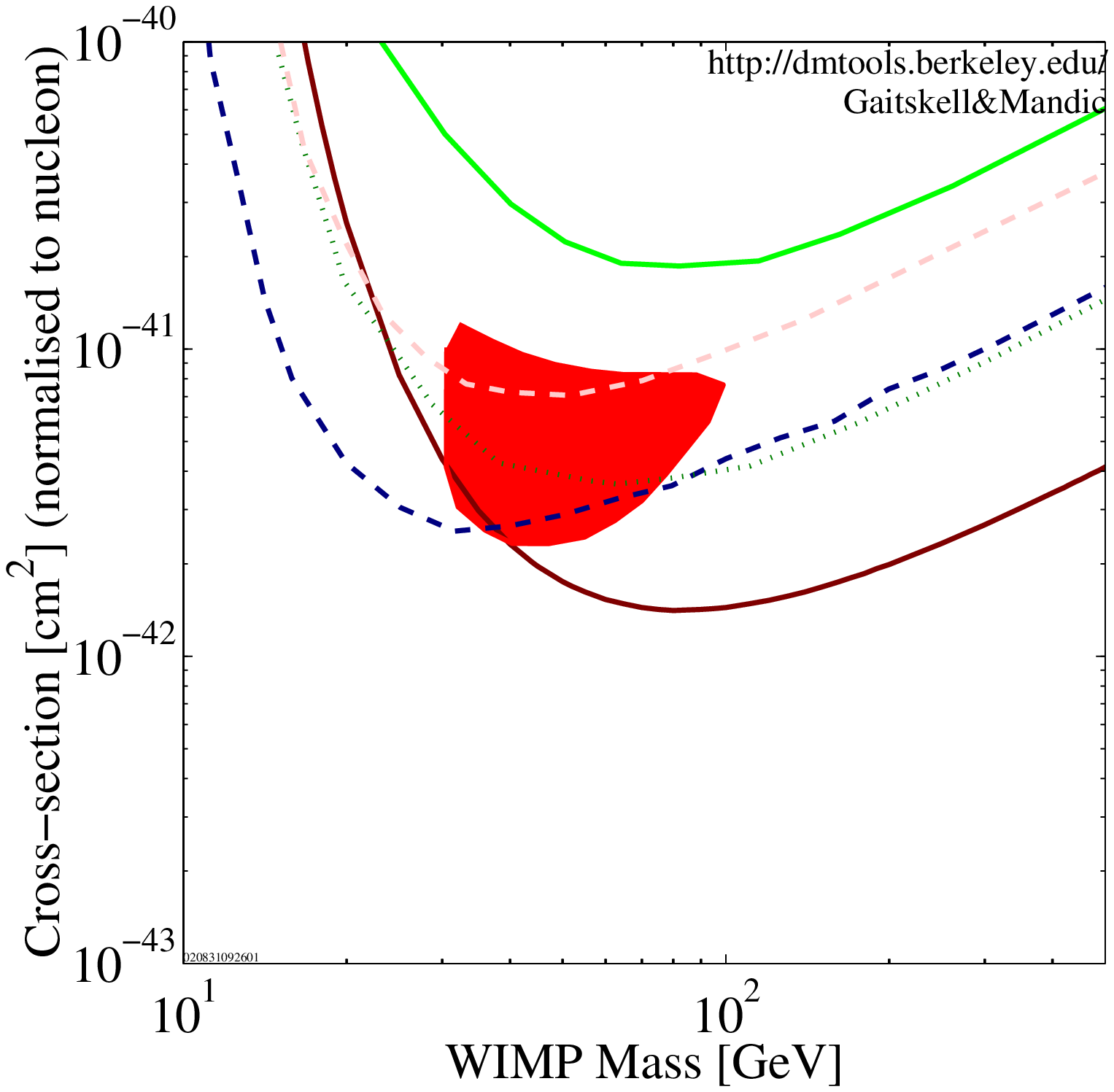}\hspace{1mm}
\epsfysize=3cm
\epsfxsize=6.5cm
\epsfbox{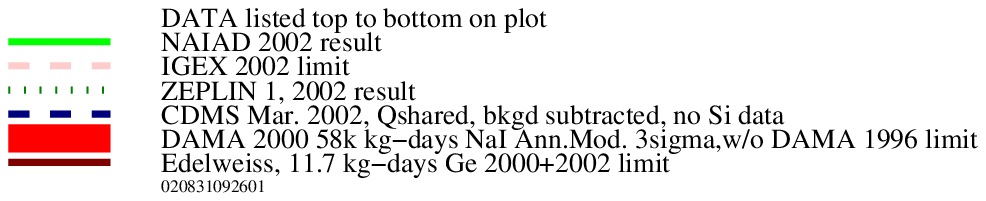}
\caption{Some selected spin-independent exclusion limits including the
DAMA evidence 
region. To the right, the figure legend is displayed.}
\label{fig5}
\end{figure*}
\par
The DAMA experiment deserves a special note at this place since it is
the only experiment which announced an evidence for WIMP
detection \cite{dama}. All other experiments gave upper limits on
WIMP-nucleon cross sections as function of WIMP mass (usually to 90\%
C.L.). The WIMP signature DAMA claims to see is the annual modulation
of count rates, see Fig.~\ref{dama}. So far, no alternative
explanation of the measured 
characteristics (consistent modulation parameters, only count rates at
threshold, 2 keV, to 6 keV visible energy show the modulation) has
been proposed. Consequently, the WIMP evidence remains essentially
unchallenged except for other experiments trying to constrain the
cross section -- mass region with upper limits. 
\par
The experiments utilizing ionisation detectors are considered to be
'classical' Germanium semiconductor experiments. However, the
implication of 'classical' meaning 'old-fashioned' must be
rejected. The collected experience in operating such detectors under
ultra-low background conditions combined with new ideas for their
design offers an interesting
future despite the fact that strong nuclear recoil discrimination
capabilities are not available. For example, the proposed large
mass experiments, MAJORANA \cite{majorana} and GENIUS \cite{genius}
(or its approved test facility, 
GTF \cite{gtf}) will reduce their background by completely new
shielding designs, 
which will then allow to reduce background further by an
efficient anti-coincidence measurement between several detector
modules. Due to the relatively large target mass, tens to hundreds of
kilograms of Germanium, they will also be able to use the annual
modulation signature as a WIMP-induced nuclear recoil
discrimination procedure.
\par
Pure cryogenic detectors like CRESST I \cite{cresst1}, Rosebud
\cite{rosebud} or the Tokyo LiF \cite{tokyo} setup do not offer an
intrinsic discrimination. They are nevertheless important to
establish low background cryogenic detector facilities or test
specifically very low mass WIMPs\footnote{which may become interesting again
as shown in \cite{silk}} due to their unprecedented low
thresholds \cite{cresst1}.  One important property of such detectors to
note is their apparently 100\% quenching factor \cite{quench},
i.e. the phonon
channel measures energy depositions from nuclear recoils like electron
recoils. This allows CRESST I to constrain a possible WIMP
signal at recoil energies as low as 600 eV (corresponding to 195 eV
for a Germanium detector threshold, 180 eV and 54 eV for a sodium
recoil and an iodine recoil respectively in a NaI crystal; for
quenching factors, see \cite{quenchge,quenchdama}). 
\par
Apart from the above mentioned 'pure' technology experiments, most
efforts on WIMP detection focus on 'mixed' detectors, see
Fig.~\ref{fig3}, offering two
instead of one information-readout channels. Common to these exciting
technologies is their strong nuclear recoil discrimination
capability. Again three alternative approaches exist: combining light
and ionisation readout - ZEPLIN II and III \cite{zeplin2}, phonon and
light readout -
CRESST II \cite{cresst2} and Rosebud and phonon and ionisation readout
- CDMS and EDELWEISS. 
\par
The family of ZEPLIN experiments (up to a proposal for ZEPLIN-MAX, a
one ton ZEPLIN II extension) uses liquid Xenon as target material, a
high mass nucleus. The ZEPLIN II and III detectors use in addition to
the liquid scintillator signal the ionisation produced in the liquid
to drift the released charge out of the liquid into the gas
phase. There they measure the secondary scintillation in the
gas either in a low drift-field mode or in a high drift-field
mode. Since the ionisation due to electron recoils is
much more efficient in liquid Xenon compared to nuclear recoils, an
event--by--event discrimination becomes possible for total target
masses of up to 30 kg (ZEPLIN II). Besides ZEPLIN, similar
projects have been proposed, XENON \cite{aprile}, or are under
construction, XMASS \cite{xmass}.
\par
The cryogenic phonon--ionisation experiments CDMS and EDELWEISS are
operative already and give the sharpest constraints so far on WIMP
spin-independent cross sections. They use Germanium and (CDMS) Silicon
semiconductor crystals to collect the phonon signal and the ionisation
signal (applying a small bias of about 6 V) simultaneously. The
quenching factor then yields two distinctive branches of
events as function of energy in both readout channels. Electron
recoils produce the same energy output in the ionisation as in the
phonon channel\footnote{After a known correction due to the Luke effect, which
describes phonon production due to drifting charges in the
crystal.}. Ionisation from nuclear recoils instead is quenched, so
that only the phonon channel shows the true deposited energy. The
collaborations encountered a problem arising from incomplete charge
collection, which can mimic a nuclear recoil (less ionisation than
expected from an electron recoil). This appears to have been solved at
least to a degree that they now give the most stringent limits of all
direct WIMP searches. In addition, the approach of the CDMS
collaboration to use Germanium and Silicon crystals in a common setup 
offers the chance to
discriminate for the first time between nuclear recoils due to
neutrons and WIMP events. 
\par
The combined light--phonon readout means to employ a scintillator
crystal, cooled to cryogenic temperatures (about 15 mK for
the CRESST setup \cite{cresst2}). Since scintillation measurements
need some light-sensitive device, such a detector module has to
involve two cryogenic detectors. One is a scintillating crystal with 
a phonon readout thermometer. The second is an extreme low threshold
detector, a thin, large area crystal (between 4 and 16 cm$^{2}$ have
been tested) in a light-tight setup. The low
threshold condition comes from the maximum light output of CaWO$_{4}$
crystals, which is at around
400nm wavelength corresponding 
to about 3 eV. Fig.~\ref{fig4} shows
one of the new CRESST II detector modules. Two main advantages
compared to the previous technologies can be stated. First, the larger
collection of different potential target materials since there are many
candidate scintillators with efficient light output at the required
low temperature. That, however, is still subject to extensive research
by the collaboration. Second, the apparent independence of surface
effects like incomplete charge collection as for the semiconductor
approach. The scintillator volume is fully active without dead-layers
and no need 
exists to define the active volume. The event-by-event discrimination
involves again the quenching factor. The phonon channel measures
deposited recoil energy whereas the light channel is quenched for
nuclear recoil events.
\par One might envision for
such a technology to apply several different target materials in
a common setup to utilize the material signature of WIMP events. The
CRESST collaboration currently prepares to upgrade their experiment
for a total target mass of 10 kg CaWO$_{4}$ scintillators. First
results using a single or two new detector modules of the type shown
in Fig.~\ref{fig4} (about 300 g CaWO$_{4}$ single crystal) are
expected soon. 
\par
\begin{figure*}[th]
\epsfxsize=9cm
\epsfbox{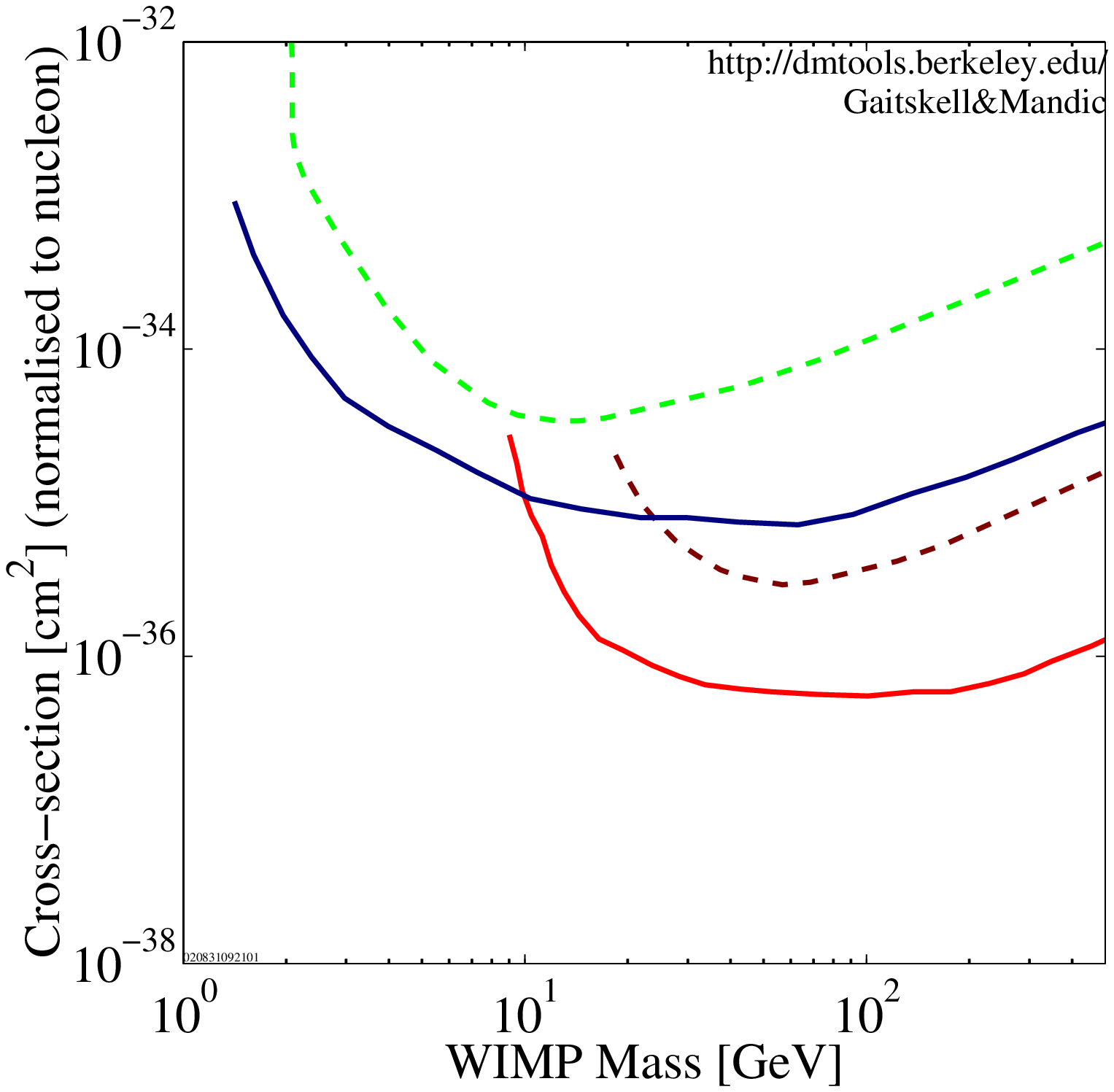}
\epsfysize=2.5cm
\epsfxsize=6.5cm
\epsfbox{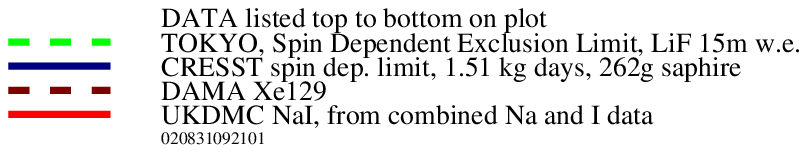}
\caption{Selected spin-dependent exclusion limits. To the right, the figure
legend is displayed.}
\label{fig6}
\end{figure*}
\par
Finally, there exist experiments and proposals which are not included
into the scheme of Fig.~\ref{fig3}. Two of them use the high
dE/dx of nuclear recoils to discriminate against background, notably
the SIMPLE \cite{simple} and the PICASSO \cite{picasso}
experiments. Metastable liquid droplets immersed in a gel expand
(explode) due to a phase transition to the gaseous phase in case a
particle or nucleus with sufficiently high energy deposition over unit
length interacts in the liquid. The threshold for such a
bubble explosion can be fine-tuned by pressure and
temperature controls. The main advantage of such integrating detectors
is that they can be tuned to be almost background-blind. They would not 
react on electron recoils and alpha-radiation events while being fully
sensitive to nuclear recoils either by fission products, neutrons or
WIMPs above a definite threshold. However, this detector type does not
deliver more energy information than the threshold. Obtaining an
energy spectrum would be a tedious procedure. One would have to vary
threshold energies step--by--step.
\par 
Then there exists another inspiring
idea, the CASPAR proposal \cite{caspar}. The idea is to use the short
range of
nuclear recoil events compared to electron recoils in order to
discriminate between the two. Small granules of scintillating crystals
are immersed in a liquid scintillator. A long range background
event then has a high probability to excite both scintillators which
would be visible by pulse shape analysis. Nuclear recoils instead
should only show the characteristic light-output of the crystalline
scintillator. 
\par
Finally, a special sort of ionisation detector shall be mentioned, the
DRIFT experiment \cite{drift}. Although it uses pure ionisation in a
gas as detection principle it deserves a special
place among the WIMP detectors. The emphasis here is on ionisation
tracks which are measured with a multi-wire proportional
chamber in a low-pressure gas. Therefore it represents the first
operating nuclear 
recoil-direction sensitive experiment. It is in fact the only
experiment (in principle) capable of observing the strong diurnal
modulation. The main drawback, however, is obvious by the term
'low-pressure', i.e. the low target mass. Nevertheless, the track
recognition also implies a strong background discrimination by recoil
range. An expansion of the experiment into the kilogram mass scale is
planned. 
\section{Conclusion}
The general aim would be to build a ton-scale experiment in
order to explore orders of magnitude lower cross sections. However, as
can be inspected in Figs.~\ref{fig5},\ref{fig6}, even with the
most promising technologies it is a long way to gain factors in
sensitivity. So far, no experiment has ever measured below the
$10^{-6}$ pb level ($10^{-42}$ cm$^{2}$) for spin-independent
interactions. It is not known, which kind of systematics will appear
for a given detector when probing an order of magnitude below. It
would certainly represent already a big leap for the field if an
experiment could reach the $10^{-7}$ pb level. The point is that
simple scaling of existing technologies will not be sufficient as
unknown or so far unstudied effects might become dominant sources of
background. Dedicated studies beforehand of thinkable extreme rare
events would certainly help to decide for a detector technology to
scale-up for a large experiment. A first study in this direction, for
example, has been undertaken by the EDELWEISS collaboration
\cite{edelweiss3}. 
\par
Besides these efforts, one should bear in mind that the DAMA evidence
still is neither excluded nor confirmed. Therefore it might
turn out as well that experiments encounter an irreducible
'background' simply being WIMPs. In that case, building an experiment
with a sensitivity between the $10^{-6}$ to $10^{-7}$ pb level would
produce in fact
a high signal-to-noise WIMP spectrum which can be studied in detail. Such
a sensitivity level is promised for the next round of experiments like
CRESST II, CDMS, EDELWEISS and the ZEPLIN family among
others. Therefore, it might be that in the near future an exciting
discovery of dark matter particles could be announced. 
\par
Inspecting Fig.~\ref{fig5}, one might question the statements from
above as the new EDELWEISS limit seems to exclude practically the whole
DAMA evidence region to 90\% C.L.\footnote{the same is true for the
most recent ZEPLIN I limit \cite{zeplinnew}.} However, this impression is
misleading. Such a figure provokes a combination of experimental
results 'by eye' but the only statistically justified combination of
experimental results in this case would be by multiplication of the
corresponding likelihood functions (see e.g. \cite{statistic}). Such a
comparison or combination of results has not been undertaken. Toy
model studies \cite{yorck}, however, suggest that such an operation
would rather lead to a new evidence region, incorporating the
EDELWEISS result as a new constraint. 
\par
In summary, the present situation for direct WIMP dark matter searches
is very promising with a lot of upcoming results from diverse detector
technologies in the near future. 

\end{document}